\documentclass[conference]{IEEEtran}

\usepackage[numbers]{natbib}
\usepackage{amsmath,amssymb,amsfonts}
\usepackage{algorithm} 
\usepackage{algorithmic}
\usepackage{graphicx}
\usepackage{textcomp}
\usepackage{xcolor}
\usepackage{url}
\usepackage{flushend}

\def\BibTeX{{\rm B\kern-.05em{\sc i\kern-.025em b}\kern-.08em
    T\kern-.1667em\lower.7ex\hbox{E}\kern-.125emX}}
    
\makeatletter
\renewcommand*{\ALG@name}{\scriptsize Procedure}
\makeatother

\begin{document}

\title{Estimating Return on Investment for GUI Test Automation Frameworks}

\author{
    \IEEEauthorblockN{\small Felix Dobslaw\IEEEauthorrefmark{1}, Robert Feldt\IEEEauthorrefmark{1}, David Micha\"elsson\IEEEauthorrefmark{1}, Patrik Haar\IEEEauthorrefmark{2}, Francisco G. de Oliveira Neto\IEEEauthorrefmark{3}, Richard Torkar\IEEEauthorrefmark{3}}
    \IEEEauthorblockA{\small \IEEEauthorrefmark{1}Dept. of Computer Science and Engineering, Chalmers University of Technology
    \\\{dobslaw, robert.feldt\}@chalmers.se}
        \IEEEauthorblockA{\small \IEEEauthorrefmark{2}Canea Partner Group}
    \IEEEauthorblockA{\small \IEEEauthorrefmark{3}Dept. of Computer Science and Engineering, University of Gothenburg
    \\\{francisco.gomes, richard.torkar\}@cse.gu.se \vspace{-.4cm}}
}

\maketitle

\thispagestyle{plain}
\pagestyle{plain}

\begin{abstract}
Automated graphical user interface (GUI) tests can reduce manual testing activities and increase test frequency. This motivates the conversion of manual test cases into automated GUI tests. However, it is not clear whether such automation is cost-effective given that GUI automation scripts add to the code base and demand maintenance as a system evolves. In this paper, we introduce a method for estimating maintenance cost and Return on Investment (ROI) for Automated GUI Testing (AGT). The method utilizes the existing source code change history and has the potential to be used for the evaluation of other testing or quality assurance automation technologies. We evaluate the method for a real-world, industrial software system and compare two fundamentally different AGT frameworks, namely Selenium and EyeAutomate, to estimate and compare their ROI. We also report on their defect-finding capabilities and usability. The quantitative data is complemented by interviews with employees at the company the study has been conducted at. The method was successfully applied, and estimated maintenance cost and ROI for both frameworks are reported. Overall, the study supports earlier results showing that implementation time is the leading cost for introducing AGT. The findings further suggest that, while EyeAutomate tests are significantly faster to implement, Selenium tests require more of a programming background but less maintenance.

\end{abstract}


\section{Introduction}

A common assumption, in test automation practice as well as in research in general, whether explicit or not, is that the more activities that can be automated the better. While manual testing is often seen as mundane, repetitive, and error-prone, automated testing can lead to lower costs, increased test frequency, earlier defect identification, and higher system quality~\cite{Rafi2012BenefitsSurvey,Alegroth2015}. While the long-term vision~\cite{bertolino2007software}, as well as shorter-term impetus of test automation, is thus often to reach full automation, this is rarely realized in practice~\cite{Berner2005ObservationsTesting,Rafi2012BenefitsSurvey}. Only a small minority (ca. 6\%) of software practitioners surveyed believed in full automation~\cite{Rafi2012BenefitsSurvey}. More detailed ways of deciding when and what to automate are needed and general and high-level guidelines have started to appear~\cite{Garousi2016WhenReview}.

One testing activity that software organizations often spend many resources on is system testing at the level of the graphical user interface (GUI). There are several challenges with automating higher-level system tests~\cite{Alegroth2015VisualLimitations}, while lower-level testing oftentimes comes with good tool support and easy automation options. Despite industrial case studies that show overall benefits and cost-effectiveness, there is thus evidence that automated GUI testing (AGT) still is relatively rare among software practitioners~\cite{liebel2013state,Alegroth2015VisualLimitations}. One reason for that is that studying the direct costs and benefits involved when automating a manual test suite does not suffice, as maintenance costs over time can be prohibitive~\cite{Alegroth2016MaintenanceTesting}. While this has been pointed out in several studies, the methodologies employed are predominantly qualitative and based on a combination of interviews, expert estimations, and opinions, rather than on direct or objective observations~\cite{kasurinen2010software,Alegroth2016MaintenanceTesting}.

The simple and basic idea of this paper is to use existing source code repositories and the change history they contain to go back in time and `\textit{replay}' the history while noting the actual development and maintenance costs of the test suite. Not only can this allow more direct measurement of the costs involved in using a particular test automation technology or tool, if done carefully, but it can also allow the comparison of multiple AGT frameworks and thus provide concrete decision-support to project managers, testers, and developers. This is in line with the trend within the empirical software engineering community of a detailed analysis of software repositories~\cite{kalliamvakou2014promises}. However, in the method proposed here, instead of `\textit{mining}' the repositories with automated analysis tools, the repository is used as a historical record and the actual code changes are `\textit{replayed}'. 

The method has similarities to the `Development Replay` approach of~\citet{hassan2006replaying} but they don't expend or measure manual work to understand what it would have been; their focus is on what a certain tool would have told developers if applied at earlier points in time. In contrast, we do expend manual work by replaying history. We will show that this not only allows us to estimate actual implementation and maintenance costs but also has the benefit of providing actual artifacts that the company can then build on. While the overall idea of the methodology is a general one, and can potentially be used to evaluate many different software engineering technologies, we here focus on automation of GUI testing in an industrial context. An industrial company in Sweden that develops a web-based software application is the specific case we use to illustrate the methodology. The specific question we helped them answer was which of two different AGT frameworks they should select and what the specific trade-offs to consider are when deciding between them. What we concretely propose and evaluate is thus a context- and project-specific way for an organization to decide between test automation technologies and how to estimate their Return on Investment (ROI).

Thus, the main contributions of this paper are twofold. First, we introduce the source code history replay method to estimate the cost as well as the Return on Investment for automated GUI testing. 
For clarity in exposition, we present the approach as a method, as we think that it has general value and can be used to evaluate also other automation technologies. Second, a case study comparing two current but conceptually different AGT frameworks, Selenium (based on access to the GUI components via their ids/names) and EyeAutomate (based on image recognition on the GUI itself), for an industrial software product is conducted. We use both cost and ROI estimation of the method in this case study and consider both implementation cost (at the point in time we replay back to) and the maintenance costs (by stepping forward up until the current time, while tracking actual work).

The remainder of this paper is structured as follows. Section \ref{background} gives some background and introduces the field of AGT along with related research. In Section \ref{method}, the step-wise replay method for ROI estimation gets introduced, and in Section \ref{case-study} the case study gets detailed. The results of the case study are presented in Section \ref{results}, and, together with insights regarding the step-wise replay, are discussed in Section \ref{discussion}. Finally, Section \ref{conclusions} summarizes the main findings and concludes the paper.

\section{Background and Related Work}
\label{background}

This section includes discussion on terminology, concepts, and practices on GUI testing, as well as the different studies reported in the literature that reveal the trade-offs with such techniques.


\subsection{GUI Testing}
GUI testing is the activity of testing a system by interacting with the graphical elements presented to the user (e.g., buttons, forms, drop-down menus)~\cite{Memon2002GuiProcess}. Consequently, GUI testing is on a higher level of abstractions if compared to, e.g., unit-level tests where knowledge of the source code is relevant to understand and maintain the test. For instance, testers can test the System-Under-Test (SUT) by entering data through point\slash click interfaces, navigating between a number of views in a predefined order or even fill in forms with dynamic visibility\slash activation constraints. Problems with the SUT or test environment should be visible via the GUI, hence leading to desired test outcomes (i.e., either pass or failure). However, \textit{undesired} GUI related test outcomes are also possible, such as the toleration of a missed button click test (\textit{false negative}, i.e., a test that passes but should fail) or the failing of a test for a correct click sequence due to, e.g., a timing issue (\textit{false positive}, i.e., a test that fails but should pass).

In order to spare human testers of the tedious, laborious and error-prone work of manually interacting with the GUI of a SUT (also referred as Manual GUI Testing--MGT), various techniques and tools that enable automated GUI testing have been proposed in research~\cite{Alegroth2017OnStudy,Grechanik2009MaintainingScripts} and practice~\cite{Alegroth2016MaintenanceTesting,Alegroth2014IndustrialLearned,Borjesson2012IndustrialAutomation}. Even though Automated GUI tests (AGT) are more reliable, reproducible, and tolerant to small GUI related changes in the SUT~\cite{Alegroth2015ConceptualizationStudy}, they demand maintenance of the test suite to match the new GUI~\cite{Alegroth2016MaintenanceTesting}. On the other hand, humans who manually test a GUI can easily adjust to a re-located, re-named, or re-sized button.

We investigate the use of AGT frameworks. Frameworks can be differentiated from tools in that they offer and require more customization, whereas tools serve a very specific need and often work right away. Tools in the sphere of AGT, such as the Application Exerciser Monkey tool for Android UI testing \cite{android}, can only offer superficial testing support, and lay therefore outside the scope of this paper.

AGT frameworks are classified into three generations according to the underlying elements of the GUI used to automate the test, such as $i$) the coordinates of the GUI (first generation), $ii$) the elements\slash objects (buttons, forms, menus, etc.) of the GUI (second-generation) and $iii$) images of the GUI (third-generation). Below, we briefly distinguish the different generations and present their trade-offs.

The first generation of AGT frameworks assists testers by creating recordings of traces of movements, mouse-clicks, and other interactions that represent use cases by tracking the coordinates of such actions~\cite{Potter1993Triggers:Access,Zettlemoyer1999IBOTS:Interface}. Such frameworks have a low technical usability threshold, but require much manual maintenance, hence being abandoned and replaced by a new generation of GUI testing frameworks.

Second-generation frameworks interact with the GUI using references to its elements such as references to Java Swing objects, or buttons of a Document Object Model (DOM). On the one hand, the test benefits from targeting defined GUI element locators such as IDs, type or labels, regardless of their layout. On the other hand, a programmer would need to know these locators and the composition of GUI elements which can hinder the implementation of the test. This generation frameworks are more stable due to their reliance on structural properties, hence being popular in industry~\cite{Ardito2018issta}. Still, they require programming knowledge from the tester. Selenium~\cite{selenium} is a well-known second-generation GUI testing framework that automates interactions with browsers and web applications.

In turn, third-generation AGT, referred to as \textit{image-based testing}, uses image recognition to navigate the GUI~\cite{Alegroth2015ConceptualizationStudy}. Therefore, use-cases are described in terms of images from the GUI itself and high-level operations (e.g., commands to click, drag or re-size the GUI) which are test implementation independent of programming language or frameworks, such as Java or Python. Such tools attempt to find the sweet spot between the technical threshold for entry and the re-usability of the use-cases due to an automated interpretation of the user sequences into emulated event-streams~\cite{Alegroth2015ConceptualizationStudy}. Sikuli~\cite{sikuli} and EyeAutomate~\cite{eyeautomate} are two examples of third-generation frameworks evaluated in research~\cite{Ardito2019ease} and practice~\cite{Alegroth2017OnStudy}.

Introducing automated tests can significantly increase the testing frequency to find errors early (\textit{fail fast, fail cheap}), allowing quality assurance even for short deployment cycles. However, there is a cost both in implementing and maintaining the set of automated GUI tests that  must be considered when evaluating the adoption of an AGT framework in a software project. 

\subsection{Empirical evaluation of GUI testing}
\label{related_work}

Automated GUI Testing has been widely investigated in the literature. Recent investigations focus both on understanding the complementary aspects between second and third-generation frameworks~\cite{Ardito2019ease,Ardito2018issta,Alegroth2017OnStudy}, and the automation advantages over manual GUI testing~\cite{Alegroth2015ConceptualizationStudy}. Several frameworks have been proposed for both the element-based (GUITAR~\cite{nguyen2014guitar}, Selenium, DART~\cite{Memon2005AutomatingSoftware}, Espresso~\cite{espresso}, etc.) and visual-based (Sikuli, JAutomate~\cite{jautomate}, EyeAutomate) generations. The frameworks differentiate between scripting and recording tests~\cite{Leotta2013Capture-replayEvolution}, as well as performance tests \cite{Adamoli2011AutomatedTesting}.

Different frameworks have been applied in a variety of test activities including repairs of regression test suites (e.g. using GUITAR~\cite{Memon2008AutomaticallyTesting}), daily/nightly GUI test automation~\cite{Memon2005AutomatingSoftware}, web testing~\cite{Ricca2002TestingApplications,Marchetto2008AApplications}, Android testing~\cite{Ardito2019ease,Ardito2018issta}, and the evaluation of new GUI testing techniques itself~\cite{Michail2005HelpingApplications}. Moreover, second-generation frameworks have been particularly useful for agile methodologies~\cite{Holmes2006AutomatingSelenium}, and have shown benefits over manual GUI testing~\cite{Alegroth2015ConceptualizationStudy,Alegroth2016MaintenanceTesting}. Several such studies have been conducted in the industry.

In a comparative study,~\citet{Ardito2019ease} compare Espresso (second-generation) and EyeAutomate (third-generation) contrasting both AGT frameworks in terms of productivity (total number of created tests) and quality of the created tests (ratio between correct and written tests). Even though there was not a significant difference in both the framework's learnability and productivity, the authors found that EyeAutomate enables the creation of \`{}better\`{} tests. The authors also concluded that practitioners had a marginal preference to third-generation frameworks given the intuitiveness and ease of use in creating test scripts through screen captures.

A similar investigation has been conducted by \citet{Alegroth2015VisualLimitations} where AGT (third-generation) is compared to manual testing in terms of return on investment in two cases. The authors reveal that the ROI of transitioning manually written tests into visual GUI testing scripts would be positive within one month after completion. However, the study includes many limitations in their ROI model, since the cases had different contexts, different number of script developers, and did not consider, explicitly, the costs of maintaining the GUI testing scripts. The use of ROI to compare manual and visual GUI testing is refined by \citet{Alegroth2016MaintenanceTesting}, revealing that it takes longer to reach positive ROI after automation for companies not highly invested in manual testing. In other words, for small projects not supposed to scale in size, the costs to introduce visual GUI testing do not outweigh the costs of adopting it.

To the best of our knowledge, no general model for the assessment of ROI when evaluating AGT has been proposed. The explicit formalization of source-replay for efficient utilization of existing software adds to the contribution. Despite the several attempts of evaluating AGT in terms of ROI, researchers still limit the conclusions based on the constraints introduced to their ROI model, such as using specific cost constructs (e.g., time taken to create the tests, execution time) or introducing weak assumptions, such as similar costs for test maintenance and test execution.
Our study advances on existing investigations by proposing a model for ROI that is versatile and can be used with different cost constructs (timeline of the project, execution time, etc.). We evaluate our proposed model using similar cost constructs used in literature (i.e., maintenance in terms of time), but instead, focusing on the comparison of different generation AGT frameworks within the same context. We further argue that its construction can be applied to the introduction of other cross-cutting software quality concerns, e.g., within the broader scope of testing, but also for reliability and security enhancements.

\section{Manual Step-wise Replay Testing}
\label{method}
Most software projects today use version control software (such as Git\cite{git}) for source code management. The source history of software documents its evolution over time and simplifies collaboration as well as automation. We propose a method to estimate the cost, and consequentially the ROI, for introducing AGT into existing software projects. The method makes use of the software's source history, including compulsory manual steps for source-change management.

\begin{figure}
\centering
\includegraphics[width=.7\columnwidth]{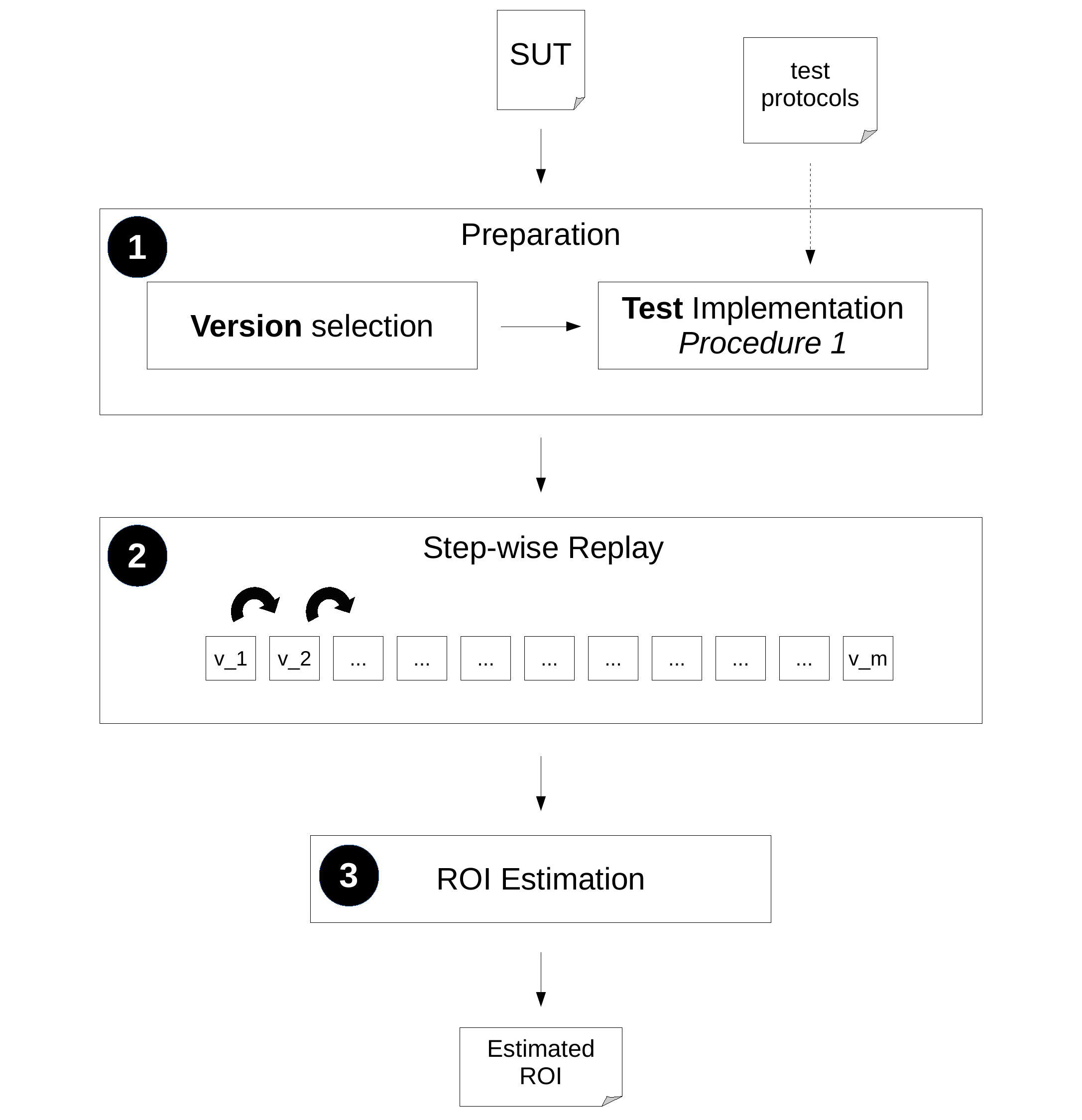}
\vspace{-.3cm}
\caption{The proposed manual step-wise replay method. 1. Software versions for the step-wise replay are shortlisted and test protocols (test cases and related documentation) implemented as automated GUI tests. 2. Manual step-wise replay with time measurements and logging for maintenance is sequentially and chronologically conducted for the selected versions. 3. The ROI gets estimated by comparing the manual testing costs to the implementation and maintenance costs for the AGT frameworks and associated GUI tests.}
\label{fig:roi-method}
\end{figure}

We summarize the three phases of step-wise replay illustrated in Fig. \ref{fig:roi-method} and detailed further below. In phase one, two selection activities for the step-wise cost assessment are conducted. First, the sequence of software versions to be investigated is selected from source-control. Second, orthogonally to the versions, representative test protocols are selected and implemented into test scripts for automated execution. Here, a test protocol is the description of a test case that documents the steps required to test the expected behavior of the SUT. Phase two is the step-wise replay procedure, where, for each selected version, in sequence, each automated test script gets executed and potentially upcoming failures are being handled by the tester directly to ensure that the tests pass. All maintenance costs get categorized and measured based on the necessary maintenance actions performed by the tester, e.g., changing the test code, handling found bugs (more on that further below), or simply not doing anything (cheap maintenance). This is repeated until the last version has been reached and all test scripts pass. In phase three, the ROI is estimated based on the cost measurements from phase two.

Conducting a replay of overhauled historical code eliminates the potential of introducing a bias into the ongoing product development, since developers had not been aware of the study while developing. In other words, the developers working on the latest code branch are not hindered by testers studying the maintenance of the SUT. Because the replay bases on the readily available source history, long development periods can be replayed within much shorter time, i.e. years of development can be replayed within days of investigation. The estimation is thereby not dependent on the current development progress of the product. The method, as presented, measures actual costs in terms of \textit{time spent} on activities, also as the basis for the ROI. Time spent can trivially be translated into, e.g., the actual costs per hour. However, other cost constructs can be added to that and explored in future work, such as costs for the purchase of the AGT software itself, the effects of increased test frequency on quality, etc.

There are two main assumptions with our approach: First, we limit our scope to parts of the SUT where use-cases are stable throughout the time-span of their considered source code history. In other words, we assume that the SUT has not evolved away from its original intent/specification. Second, we assume that the software on the relevant branch of the source-management system builds and deploys without errors. Those assumptions are relevant to avoid noise to the ROI estimation, such as writing test scripts that do not pass due to a broken build or an unstable version of a feature, as opposed to the cost-effectiveness of the GUI testing technique itself. We argue that those assumptions are reasonable given that global commits require unit\slash regression tests to pass first according to good industrial practice. 

An underlying assumption is, of course, also that the implementation and maintenance costs as performed now, after the fact, are similar to what the costs would have been if they had been done at the earlier points in time. There are threats to this assumption, in particular, if the investigation was done by developers that built the system. Learning effects, i.e. them knowing what the system eventually turned out to be, could make their work more efficient in an unwarranted way. Conversely, the effects of forgetting the earlier design and state of the system might have the opposite effect. We argue that this can be overcome in practice by having more independent developers or testers do the actual replay work. In fact, in the case study, they were even external to the company at the time of doing the work.

\subsection{Phase one: Preparation}
The creation of the AGT scripts is precluded by a selection of a subset of the existing test-protocols. This is because the implementation costs may be prohibitive for the entire manual test-suite, defeating the purpose of a cost-effective ROI estimation. Only after the step-wise replay, once the ROI has been estimated based on the sub-selection and in case the decision was made to go ahead with AGT for the entire SUT, are the remaining test-protocols implemented. The sub-selection should consider a good mix of randomization and expert-knowledge.

To clarify the methodology and its steps, a number of variables will be introduced below. For convenience, find all their definitions in Table \ref{tab:variables}. Let $c = c_1, \dots, c_n$ be the chronological commit history of the SUT's version-control system (or branch we are focused on), with $c_1$ and $c_n$ being the first and latest commits, respectively. Since $n$ could be a large number and each commit considered for the estimation requires time, we select a chronological sub-sequence of $c$, namely $v = v_1, \dots, v_m$ with $m \leq n$ for the investigation. We call $v_i$ with $1\leq i\leq m$ \textit{version} $i$ of the software.
\begin{table}
    \scriptsize
    \centering
    \caption{Step-wise replay testing variable declarations.}
    \begin{tabular}{ll}
    \hline
    \textbf{Identifier} & \textbf{Description} \\
    \hline
     $c=c_1,...,c_n$ & SUT commit history \\
     $v=v_1,...,v_m$ & SUT version history for study \\
     $T$& All SUT test protocols\\
     $T_+ \subseteq T$& Shortlisted SUT test protocols\\
     $T_X$& SUT test protocol implementations\\
     $A$& Set of investigated AGT frameworks\\
     $\tau_t$& Time to manually conduct the test $t\in T$\\
     $\tau_{t_\alpha}=\tau_{t_\alpha, 0}$& Implementation time of $t\in T$ using $\alpha \in A$\\
     $\tau_{t_\alpha, i}$& Maintenance time for $t\in T$ using $\alpha \in A$ in version $v_i$\\
     $\tau_{T_+}$& Total manual testing time for $T_+$\\
     \hline
    \end{tabular}
    \label{tab:variables}
    \vspace{-.4cm}
\end{table}
The strategy for selecting $v$ is project specific and may be chosen arbitrarily but with care. It shall be sensible to key factors such as length of the considered time interval, frequency of commits, code-churn, the total length of history, or the required reliability of the estimate. Selection strategies may be based on time intervals, such as weeks, or after reaching a code-churn of $x$ lines in-between commits has been reached.

Procedure \ref{alg:test-cases} details the implementation of test protocols into automated GUI test scripts with the version $v$ vector as input. The sub-selection of test protocols for implementation $T_{+}$ happens in line \ref{alg:test-case-selection}. This selection shall consider properties such as relevance and code coverage. For instance, in our case study (Section ~\ref{case-study}), the developers suggested a few important test protocols, while others were randomly selected.

Next, a tester\slash researcher manually executes each test $t\in T_{+}$ and records the time spent $\tau_{t}$ (line \ref{alg:manual-test}), representing the costs for manual GUI testing. For each framework $\alpha \in A$, a tester implements an automated test script $t_\alpha$ and records the implementation time $\tau_{t_\alpha}$ (line \ref{alg:automated-test}). With the baseline measurements of time spent on manual testing and implementation times in place, the step-wise replay (phase two) can start.

\begin{algorithm}
        \scriptsize
        \caption{\scriptsize Test Implementation}
        \begin{algorithmic}[1]
        \label{alg:test-cases}
            \REQUIRE SUT version sequence $v$, set of test-protocols $T$, set of AGT frameworks $A$
            \STATE select representative sub-set $T_{+} \subset T$ \label{alg:test-case-selection} considering $v$
            \FOR{$t \in T_{+}$}
                \STATE manually conduct $t$ on $v_1$ and record the time $\tau_t$ \label{alg:manual-test}
                \FOR{$\alpha \in A$}
                    \STATE implement automated test $t_\alpha$ that covers $t$ and record implementation time $\tau_{t_\alpha}$ \label{alg:automated-test}
                \ENDFOR
            \ENDFOR
            \RETURN automated tests $t_\alpha$, manual testing times $\tau_t$, implementation times $\tau_{t_{\alpha}}$
        \end{algorithmic}
\end{algorithm}
\vspace{-.4cm}

\subsection{Phase two: Manual step-wise replay}
The step-wise replay is described in Procedure \ref{alg:hist-testing}. Let the set of automated tests $T_X$ be defined as $T_X = \{~t_\alpha~|~\forall t \in T_{+},~\forall \alpha \in A\}$. Step by step, for all versions $v_i$ with $i \in \{1,...,m\}$, all tests in $T_X$ are executed (line \ref{alg:line:pass-fail}). In case the system fails, maintenance activities are carried out and maintenance time $\tau_{t_{\alpha}, i}$ is recorded and summed up for that step until all failures are resolved (lines \ref{alg:line:failure-start}-\ref{alg:line:failure-end}). Before going over from step $i$ to $i+1$, the tester should document the faults found as well as the fixes\slash workarounds created in order to stop the fault from triggering in subsequent steps. If a build breaks, i.e., the code does not compile, it gets corrected, re-run, and updated in the AGT test-suite on-wards. Occasional crashes of the SUT or AGT framework are treated by re-running the tests.

\begin{algorithm}
        \scriptsize
        \caption{\scriptsize Step-wise Replay}
        \begin{algorithmic}[1]
        \label{alg:hist-testing}
            \REQUIRE SUT version sequence $v_1,...,v_m$, set of automated tests $T_X$
            \FOR{$i = 1,...,m$}
                \STATE checkout $v_i$ from the version control system
                \STATE make sure the software builds correctly
                \FOR{$t_{\alpha} \in T_{X}$}
                    \STATE run $t_\alpha$ and record outcome (pass/fail) \label{alg:line:pass-fail}
                    \IF{\texttt{test failed}}
                        \STATE record maintenance time $\tau_{t_\alpha, i}$ for all below \label{alg:line:failure-start}
                        \IF{\texttt{bug}}
                            \STATE record bug 
                            \STATE create fix\slash workaround in SUT
                        \ELSIF{\texttt{breaks}}
                            \STATE correct $t_\alpha$ to pass test and update in $T_X$ \label{alg:line:failure-end}
                        \ENDIF
                    \STATE re-run $t_\alpha$
                    \ENDIF
                \ENDFOR
            \ENDFOR
            \RETURN maintenance cost $\tau_{t_\alpha, i}$
        \end{algorithmic}
    \end{algorithm}
\vspace{-.4cm}

\subsection{Phase three: ROI Estimation}
\label{roi_assessment_method}
The maintenance costs $\tau_{t_{\alpha, i}}$ can be plotted over the sequence of versions for each AGT framework $\alpha \in A$. A historical view of the maintenance demand over time is obtained. For reasons of conformity, and without loss of generality, we define the implementation cost of a test-case $t_\alpha$ from Procedure \ref{alg:test-case-selection} as $\tau_{t_{\alpha, 0}} = \tau_{t_{\alpha}}$. In order to estimate the ROI, we compare $\tau_{t_{\alpha, 0}}$ to the total cost of executing the manual tests for each test session $\tau_{T_+} = \sum_{t \in T_+}\tau_t$. If tests are done periodically with approximately constant time-expenditure, this cost can be represented by a linear cumulative model, $\epsilon_{T_+} = c \dot \tau_{T_+}$, with a constant growth factor $c$ over time that depends on the frequency of manual testing. 

If manual testing is conducted infrequently, the time until positive ROI is achieved may be longer than the study time, i.e. $\epsilon_{T_+}$ grows only gradually and the cumulative models for automated and manual testing do not intersect. In those cases, a model for the prediction of the estimate for the AGT cumulative cost $\epsilon_{\alpha, T_+}$ can be fitted for each framework $\alpha \in A$ in order to extrapolate the expected long term costs.
Due to the initial cost of implementation for test cases in $T_X$, and the gradual familiarization with the automation frameworks\slash code, may the relationship well be logarithmic, as suggested by the ROI model presented by~\citet{Alegroth2015ConceptualizationStudy}. A logarithmic model implies a combination of several factors that come together so that most costs are the initial ones while later maintenance and changes take less and less time as the test suite matures, the tested system more stable, and testers become more and more familiar with the framework and the test scripts.


\section{Case-study}
\label{case-study}
We evaluated the method presented above in a case study to find out whether the proposed step-wise replay can create concrete value, primarily regarding maintenance cost estimation for an industrial software system of reasonable complexity. This question could otherwise not be answered in a simple experimental setup. Maintenance costs could be estimated prospectively, i.e. going forward in time, but this would be very hard to `speed up` i.e. we would need to do this probably during several months to reach reasonable confidence levels in our cost and ROI estimates. By replaying time instead, we can support decisions about test automation much more quickly, while still basing cost estimates on actual and project-relevant specifics, i.e. the actual system being tested and the actual test cases it contains.

The on-premise web-based business management tool CANEA ONE, which is used by more than 200 organizations worldwide, serves as the SUT for this study. CANEA ONE is composed of many different programming languages, with the majority of code being written in C\#, Type-/JavaScript, and HTML. The C\# code-base alone contains more than 250.000 lines of code, and the web-application comprises of more than 100 unique pages. The source code is automatically tested through roughly 2700 unit tests. The only UI testing used in CANEA ONE is load and performance testing through e.g. JMeter~\cite{jmeter}. Most of the UI testing is done manually by a team of testers.
CANEA ONE has 20 test-protocols that are conducted four times a year. However, the intention with introducing AGT is not to run four times a year in future, but as often as possible, preferably even for automation in a continuous integration workflow, and regression testing on a daily basis, in order to catch faults immediately.

\subsection{AGT frameworks}
We evaluate the introduced method with the second-generation AGT framework Selenium and the third-generation AGT framework EyeAutomate. To our knowledge, only few studies comparing second and third-generation AGT's have been published so far \cite{Leotta2014VisualStudy, Alegroth2015VisualLimitations, Ardito2019ease}, and none of those present a ROI estimation method for the introduction of AGT.

Selenium is an element-based test automation framework that emulates a web browser and verifies the functionality of a web application through GUI level tests. Selenium tests can be created in two different ways: through the use of \textit{Selenium IDE} or by using \textit{Selenium WebDriver}. In this study, we use Selenium WebDriver due to its lower maintenance cost through the creation of scripts using language bindings \cite{Leotta2013Capture-replayEvolution}. We adhere to the \textit{PageObject Pattern} in our tests, separating GUI logic from business logic to reduce maintenance time and increase reusability \cite{Leotta2013ImprovingStudy}. Selenium is open-source software released under the Apache 2.0 License.

\begin{figure}
    \centering
    \includegraphics[width=.7\columnwidth]{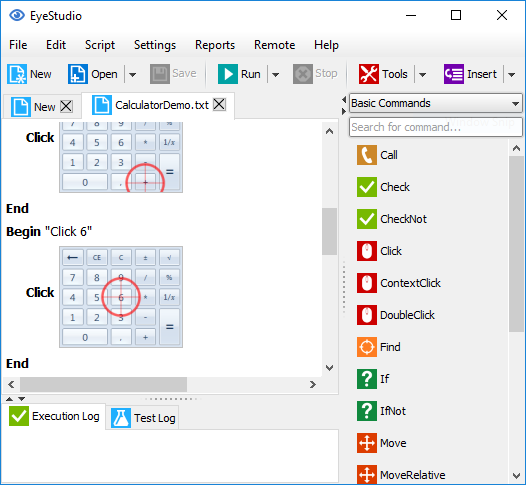}
    \caption{A screenshot of the interface EyeAutomate offers to the user. This is in stark contrast to the API level user interface Selenium offers to the user by its WebDriver.}
    \label{fig:eyeAutomate}
    \vspace{-.5cm}
\end{figure}

EyeAutomate is an image-based AGT framework that runs dedicated EyeAutomate test scripts. It was based on the earlier JAutomate system~\cite{Alegroth2013JAutomate:Automation} and has built-in support for image-recognition and uses customizable commands (to match and take actions) that the users further can extend. EyeAutomate generates a report after test runs with screenshots for all executed steps of the script for quick problem localization. The EyeAutomate scripts in this study are written using EyeStudio which offers a '\textit{What You See Is What You Get}'-like interface for script creation (see screenshot in Fig.~\ref{fig:eyeAutomate}). EyeAutomate is commercial software, but a free version, with limited features, exists.

\subsection{Test Implementation}
Six of the 20, by CANEA called \textit{critical}, test protocols were selected for implementation. Each test protocol carefully specifies a test scenario containing written instructions and details about the expected behavior. Test cases 1-3 were chosen by employees at CANEA with the intent to optimize for broad system coverage, while test cases 4-6 were selected randomly among the remaining 17 test cases. All implementation, testing, and maintenance, were conducted according to the step-wise replay method described above.


None of the authors had been using nor been involved in the development of CANEA ONE before the study and are therefore not biased by knowledge about prior or current GUI versions of the SUT.

\subsection{Version Selection}
The source-code history for CANEA ONE, which spans multiple years, has been available for the study. For practical reasons we decided the case study to be based on weekly versions from the active development branch over a total of one year, ending a week before the start of the case study. The choice of a time-based version selection is simple and straight forward, following Occam's razor, limiting bias introduced by more complex\slash subjective metrics. To control for the sensitivity of the sampling frequency, a version for each day of the week before and after that time-window was included in correct chronological order, making up for a total of 66 investigated versions. Selecting two weeks far apart might reveal non-static behavior in how the company worked on the code and\slash or tests. We worked with the development branch in order not to restrict ourselves to release candidates, which would likely have masked bugs being fixed in-between releases. Working with the active development branch also highlights the potential for AGT for regression tests.

\subsection{Step-wise replay phase}
We followed the manual step-wise replay as introduced in Procedure \ref{alg:hist-testing}, investigating the investment as the cumulative testing effort over time to identify a good fit and justify the model choice by reporting and visualizing the modeling error. The base for return on investment in our study is the time spent on manual testing compared to the time spent implementing and maintaining an automated test suite.

\subsection{Interviews}
In order to get feedback on the usability aspects of AGT, we conducted semi-structured interviews with employees at CANEA who work with the SUT. Of the six people interviewed, five were developers and one was a tester, in order to get varied opinions. The interviewees were all rather junior, with 0-10 years of working experience. None had any experience with EyeAutomate, but all the developers had experience in C\# and Visual Studio, which were used for the Selenium tests. The developers were not involved in the case study and no prior knowledge on AGT was expected. Hence, they all received an individual walkthrough with a single constant assignment for both AGT frameworks, alternating the order of frameworks to begin with, to mitigate bias. The time spent on each assignment was limited to 30 minutes; interviewees getting close to, or going over, the time limit, got assistance from a researcher. Only after this detailed introduction to each of the testing frameworks the semi-structured interview was performed. The interview focused on the interviewee's perception of the frameworks with open questions and their assessment regarding the framework's potential for CANEA ONE. The results of the interviews are discussed in the Discussion section below.

\section{Results and Analysis}
\label{results}

\begin{table*}
    \scriptsize
	\centering
	\caption{The one-time implementation times for all six test protocols.}
	\begin{tabular}{|l|r|r|r|r|r|r|r|r|}
	\hline
	\textbf{Impl. Time (min.)} & T1 & T2 & T3 & T4 & T5 & T6 & Total & $\mu \pm \sigma$\\
	\hline
	Selenium&695.8&53.35&419.37&398.33&512.72&205.33&2284.9&$380.82 \pm 226.47$\\
	EyeAutomate&346.3&19.82&127.68&296.15&183.9&220.52&1194.37&$199.06 \pm 117.49$\\
	\hline
	\end{tabular}
	\label{tab:implementation}
	\vspace{-.3cm}
\end{table*}

\begin{table*}
\scriptsize
\centering
\caption{The Maintenance statistics from the step-wise replay including total and average maintenance-times per test-protocol.}
\begin{tabular}{|l|r|r|r|r|r|r|}
	\hline
	&\multicolumn{2}{c|}{Total Time (min.)}&\multicolumn{2}{c|}{Occurrences}&\multicolumn{2}{c|}{$\mu \pm \sigma \texttt{ (min.)}$}\\
	\textbf{Maintenance}&S&EA&S&EA&S&EA\\
	\hline
	Analysis broken tests&91.25&67.35&19&22&$4.8\pm6.35$&$3.06\pm2.65$\\
	Repairing broken tests&247.18&570.72&19&22&$13.01\pm14.6$&$25.94\pm26.83$\\
	Handling found bugs&36.35&30.45&24&30&$1.51\pm4.41$&$1.01\pm2.97$\\
	Handling false negatives&56.65&10.23&2&2&$28.32\pm35.11$&$5.12\pm4.64$\\
	Handling crashes&36.1&4.05&4&2&$9.02\pm8.24$&$2.02\pm1.8$\\
	\hline
	Total&467.53&682.8&28\slash 65&26\slash 65&$7.2\pm13.74$&$10.5\pm7.78$\\
	\hline
\end{tabular}
\label{tab:maintenance}
\vspace{-.4cm}
\end{table*}

Table \ref{tab:implementation} contains the initial implementation costs, in terms of time, for both frameworks. In total, the entire EyeAutomate implementation took ca. 20 working-hours (1194 minutes, per the table), while the Selenium implementation took ca. 38 working-hours (2285 minutes); almost twice as long. Table \ref{tab:maintenance} lists statistics about the spread of time for different maintenance activities. For both frameworks, the majority of time was spent repairing broken tests. This is good for our estimation of ROI since this is likely a cost that would be a real maintenance cost if using the studied frameworks. The occurrence column contains the total number of occurrences per category. More than one occurrence is possible for a single version. Selenium could reveal 24 bugs during the study, EyeAutomate 30 bugs (25\% more). Note that even though these found bugs likely introduced additional costs during replay and would have cost even more time to fix, they are also likely to have positively affected quality since they represent bugs that the manual testing hadn't uncovered. While it can be debated if all of these costs should be counted as maintenance costs, in particular, the handling of found bugs might be seen as more of a development cost, we argue that fixing bugs is needed to maintain the value of the test suite. If a test case reveals a bug that is not fixed it is less likely to uncover other bugs. So to maintain the usefulness of an automated test suite, we likely will need to fix bugs. Thus, we have included this cost in maintenance costs. We do argue, though, that this means the maintenance costs for the automated testing tools are more likely an upper bound, rather than a lower bound.

\begin{figure}
\centering
\includegraphics[width=.8\columnwidth]{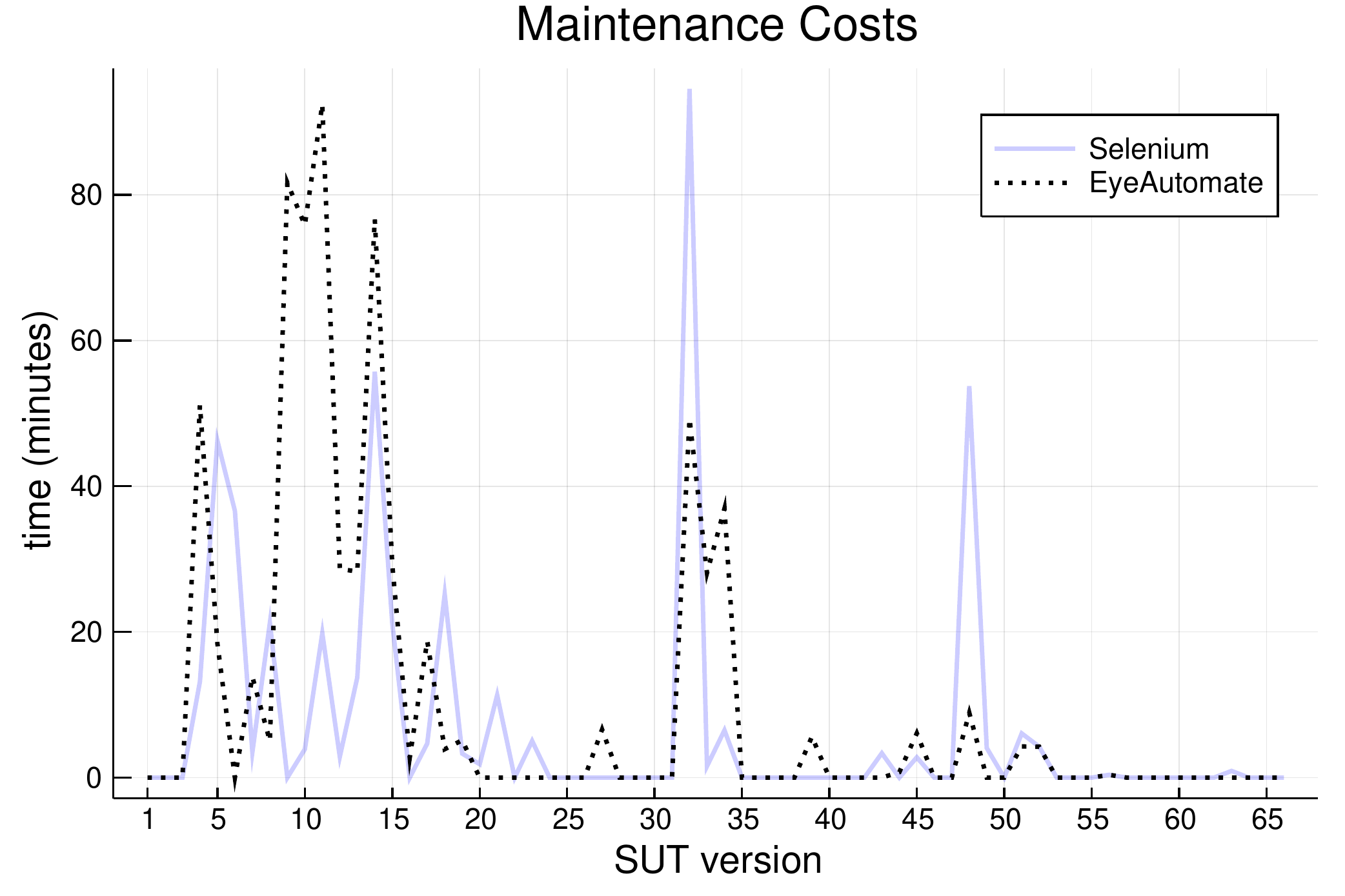}
\vspace{-.4cm}
\caption{The \textit{maintenance time per version}, here measured over one year of source control history, varies largely between the two approaches.}
\label{fig:maintanance_cost_chronologically}
\vspace{-.5cm}
\end{figure}

In turn, Fig.~\ref{fig:maintanance_cost_chronologically} shows how the maintenance cost $\tau_{t_{\alpha, i}}$ evolves over the course of the entire investigation. The spikes in the interval of versions 7 to 15 indicate larger changes in the GUI appearance.
The spike in version 31 was caused by changes to different types of GUI inputs in the SUT such as date, text, and drop-down boxes. All the spikes can thus be understood when looking at the actual changes to the SUT. The differences between the frameworks reflect actual differences in how they and the test cases are implemented and what they actually do. This further validates our methodology.

\begin{figure}[b!]
\centering
\includegraphics[width=5cm]{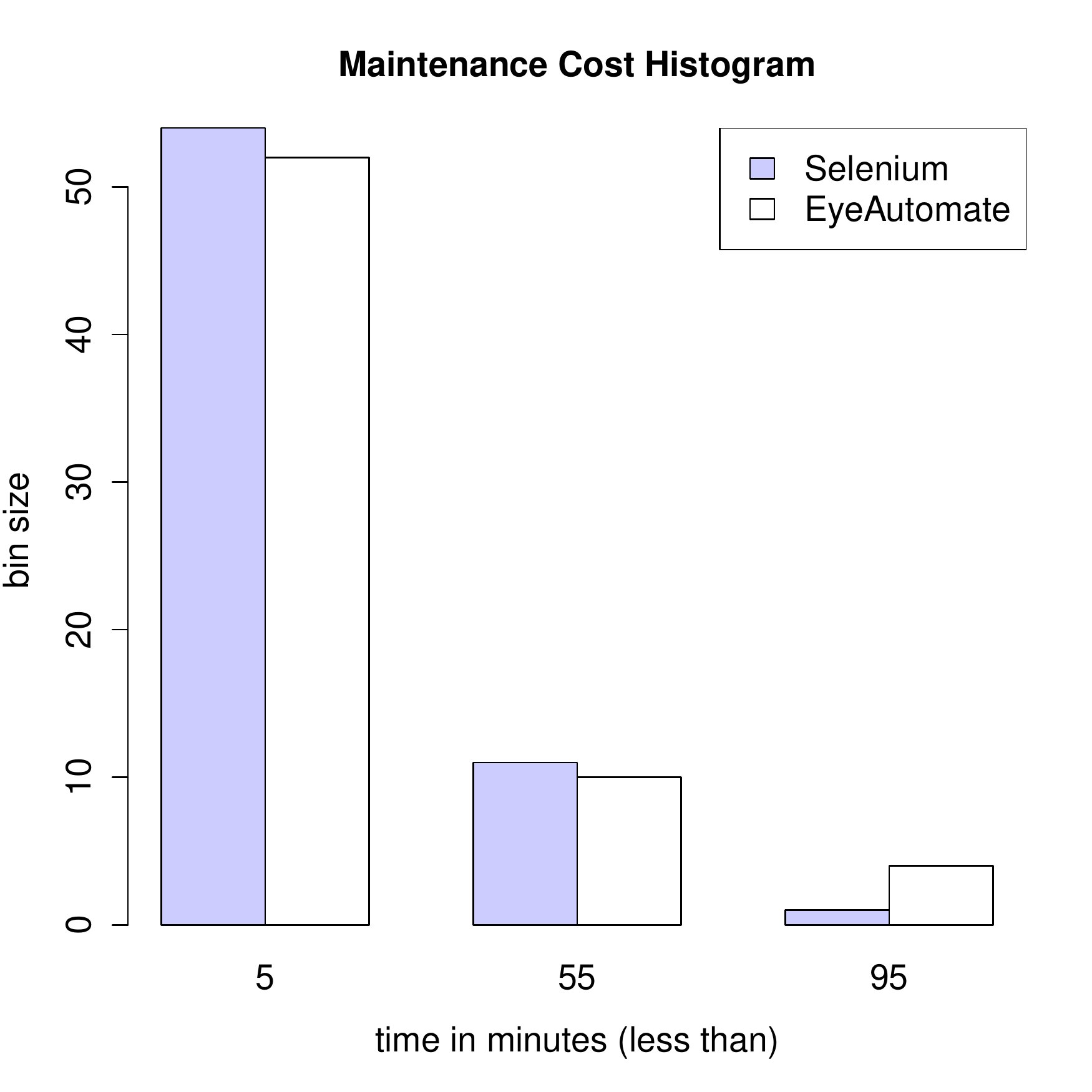}
\vspace{-.4cm}
\caption{An overall maintenance-time histogram for both AGT frameworks reveals similarities in the time-spending patterns. However, EyeAutomation test changes take an average of 32\% more time than those for Selenium, which is mostly due to a small number of versions requiring a large overhaul.}
\label{fig:maintanance_cost_histogram}
\end{figure}

Fig.~\ref{fig:maintanance_cost_histogram} presents a maintenance cost histogram comparing Selenium and EyeAutomate to see how the maintenance efforts compare in their overall distributions. The distributions are similar, but in those few situations where a change is required, EyeAutomate test cases were more likely to require longer maintenance time compared to the Selenium ones. On average, they demanded 32\% more time from the maintainer.

\begin{figure}
\includegraphics[width=\columnwidth]{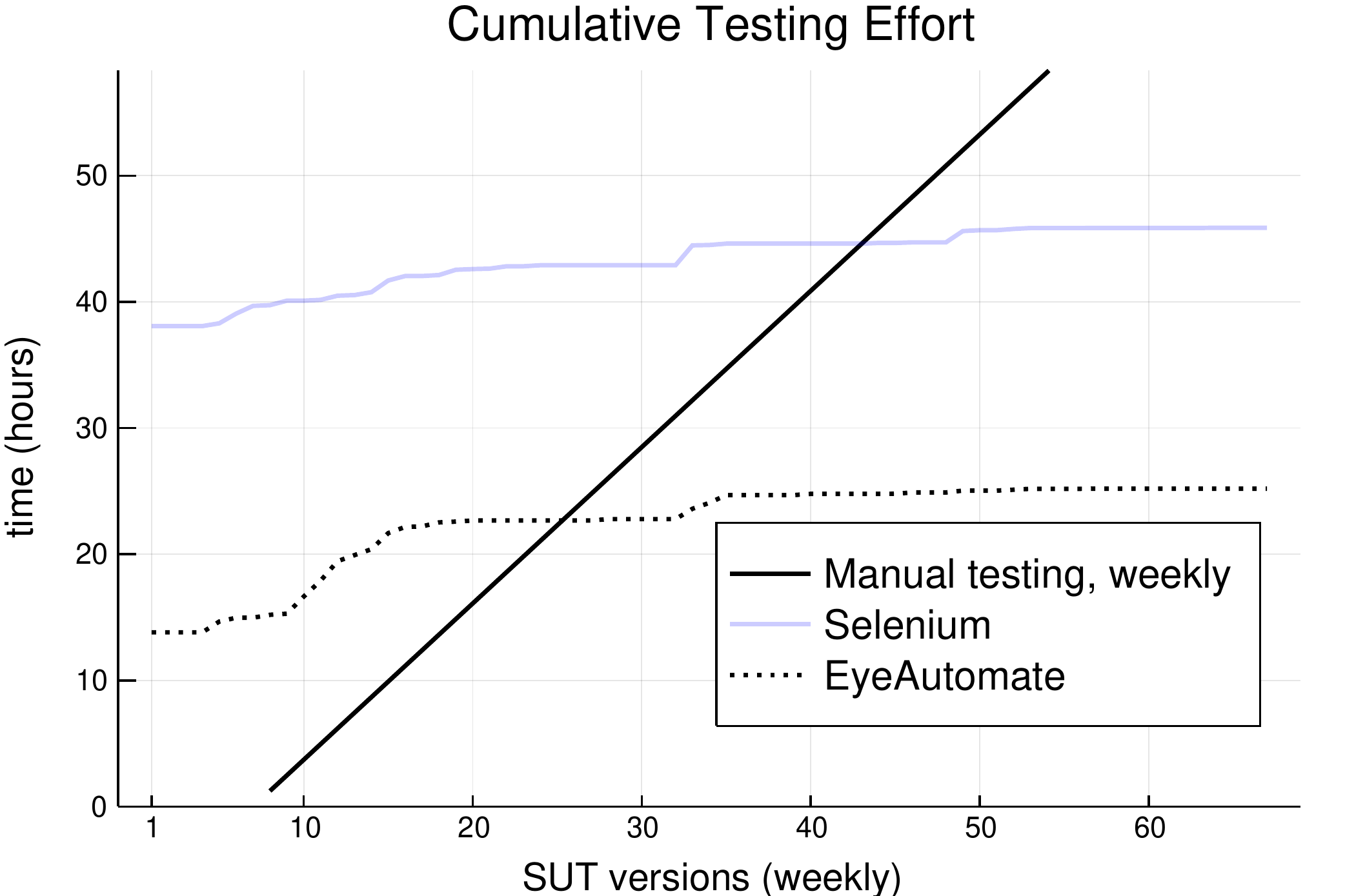}
\vspace{-.6cm}
\caption{The cumulative testing effort over time increments (SUT versions) for \textit{weekly} manual testing and the two frameworks under evaluation, Selenium, and EyeAutomate. Time increments 1-7 and 60-66 are in days, increments 8-59 are in weeks.}
\label{fig:maintenance-cumulative}
\end{figure}

In order to gain a better understanding of the maintenance demand over time, we investigated the cumulative testing effort for both frameworks over the versions based on maintenance cost $\tau_{t_{\alpha, i}}$, as described in Section \ref{method}. Fig. \ref{fig:maintenance-cumulative} illustrates the empirical cumulative testing effort over time, including the initial implementation effort $\tau_{t_{\alpha, 0}}$ for creating the automated tests. Alongside the two models, a manual testing cost, assuming weekly manual testing, is plotted as the reference for the calculation of the ROI. The ROI's for each framework, i.e. the number of versions one needs to run the tests until the costs reach the same level as would have been reached by manual testing, can be read from Fig. \ref{fig:maintenance-cumulative}. 25 versions for EyeAutomate, and 43 for Selenium. Because the first 7 versions were sampled daily, this translates into approximately 18 weeks for EyeAutomate and 36 weeks for Selenium. The main reason that EyeAutomate has a much higher ROI is its demand for initial implementation being roughly half as long, as mentioned above.

\begin{figure}
\centering
\includegraphics[width=.7\columnwidth]{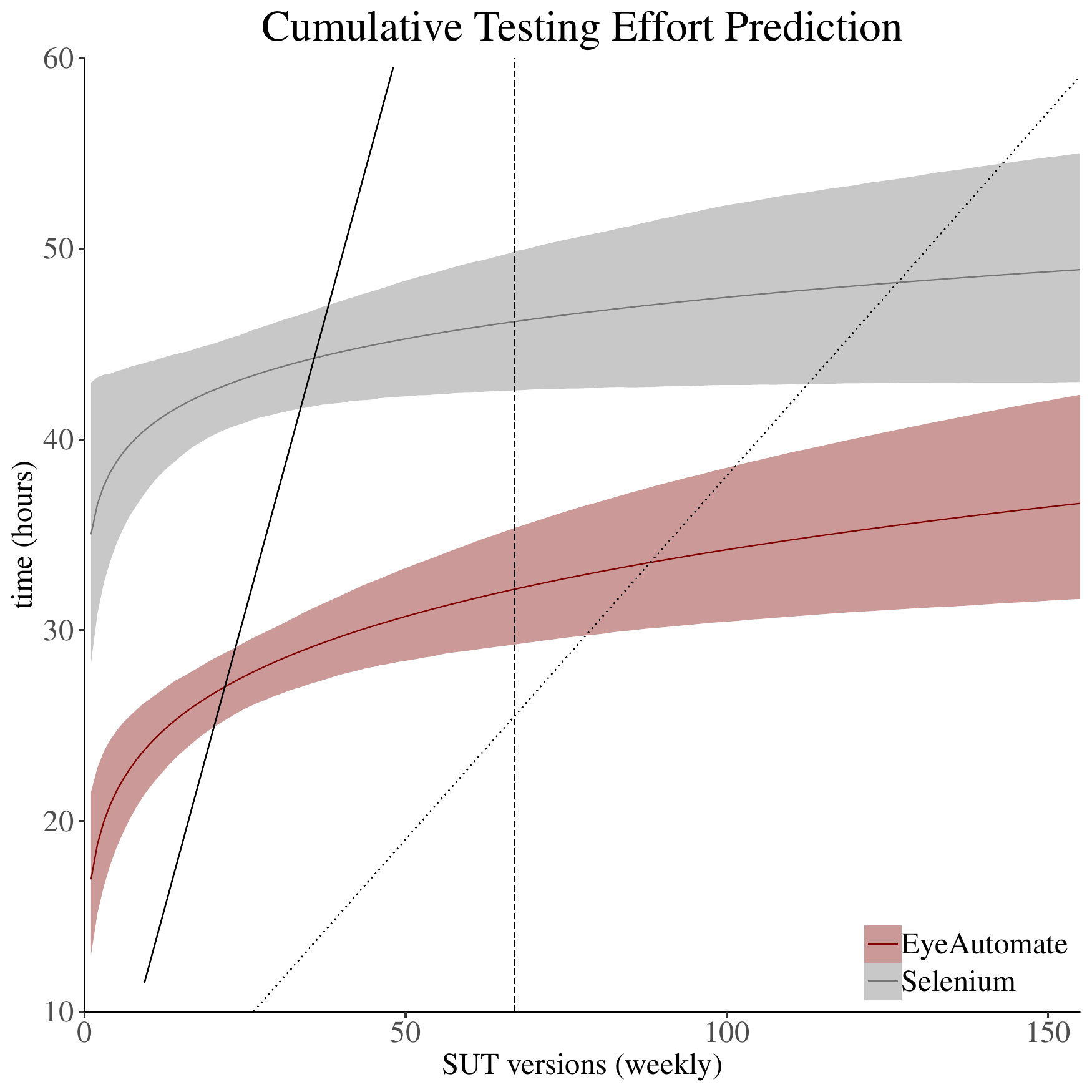} 
\vspace{-.3cm}
\caption{A model for the cumulative testing effort for the AGT frameworks can be fit in order to predict an approximate ROI. The point of (positive) ROI can be read from the graph as the intersect of the MGT and AGT functions. The solid (dotted) line models the costs for weekly (monthly) MGT.
}
\label{fig:roi_prediction}
\vspace{-.5cm}
\end{figure}

The structure of the readings leads us to further investigate $log$-based models $\epsilon_{\alpha, T_+}$ (as introduced in Section \ref{roi_assessment_method}) to predict future costs, resulting in the models presented in Fig.~\ref{fig:roi_prediction}, which are further discussed below.

\section{Discussion and Lessons Learned}
\label{discussion}

We proposed a structured process for the evaluation and introduction of automated GUI testing frameworks and evaluated it in a case study on an existing, industrial software system. This helped us identify the ROI and gave insights as to how appropriate the different frameworks are for the project's specific needs. Manual step-wise replay guides the testers, developers, and managers and gives concrete decision support for the introduction of quality ensuring measures in general, and in the scope of this paper and the conducted case study concerning automated GUI testing, in particular.

Overall, as the step-wise replay investigation is detached from the ongoing development process, independent practitioners can take on a role as the AGT specialist, i.e., working on multiple projects at the same time for reasons of cost-efficiency. After the time-consuming implementation phase, concurrent workloads can be considered. Further details and lessons learned for the CANEA ONE case study are presented below.

The maintenance demand for both AGT frameworks in Fig.~\ref{fig:maintanance_cost_chronologically} are similar in activation patterns, whereas differences in magnitude can be observed. Selenium stands for a few isolated spikes, while EyeAutomate required maintenance in bursts of multiple versions, leading to the 32\% higher maintenance costs. The similarity of activation patterns becomes apparent when looking at the histograms in Fig.~\ref{fig:maintanance_cost_histogram}. Apart from a few time-consuming maintenance operations for EyeAutomate, the histograms look very similar.

Evaluating ROI for more than one AGT framework may be prohibitive in terms of the required resources in real-world situations. Because factors such as programming competence and experience with AGT in the workplace should be weighed into the decision of AGT framework, the results here merely hint on the time required for another SUT. Applying step-wise replay, with six out of the 20 critical test-protocols investigated, required ca.~23 hours for EyeAutomate and ca.~44 hours for Selenium until ROI could be achieved, assuming weekly manual tests (see Fig.~\ref{fig:maintenance-cumulative}). Implementation costs made up for ca.~87\% (20\slash 23 hours) of the total costs for EyeAutomate, and also ca.~87\% (38\slash 44 hours) for Selenium. Assuming that the six implemented test protocols were representative, a total initial implementation time of ca.~67 hours for EyeAutomate, and ca.~127 hours for Selenium can be estimated for the entire software systems critical test-protocols initial implementation. For larger systems, with many test-protocols, the growing divide in initial time investment becomes clear.

Five out of six AGT test protocol implementations were faster for EyeAutomate than for Selenium. The exception was a date selection on a calendar view matrix, which led to many false positives in the creation phase of the automated test (T6, see Table \ref{tab:implementation}). Third-generation frameworks may have difficulties in situations where many similar alternatives for selection are visible. For instance, selecting day '23' of a month may seem similar to selecting '22' for image-based AGT. Also, those dates are, topographically on the GUI, very close to one another. Further, for a calendar with \textit{weeks per year} listed, a selection of the \textit{week} '23' field instead of the day field easily results in a false positive too. Thus, for a SUT with views containing many similar alternatives for selection, element-based AGT may, as of today, be a more sensible choice due to the localization of GUI components through, e.g., id locators. Combining different AGT frameworks where appropriate within the same SUT is certainly an alternative too. However, the development of image-recognition algorithms in later generations of EyeAutomate or other third-generation tools is going to, with time, close this.

Understanding how implementation\slash maintenance costs are affected by the complexity of the test protocols, or how they vary among the AGT frameworks, would shed some light on the error properties of AGT in general and features of the frameworks. The costs for both AGT frameworks vary largely among the six test-protocols. Using the lines of code metric for the test scripts (numbers not reported here), no non-trivial relationship could be identified other than larger test scripts require more time. It is further not possible to see which AGT framework has a generally more unified distribution of costs, both AGT's time distributions differ largely for all scripts (see Tables \ref{tab:implementation} and \ref{tab:maintenance}).

In their discussion on third-generation AGT, Alegr\'oth et al. in \cite{Alegroth2015} illustrate the ROI with the assumption that initial implementation time is steep and maintenance costs over time very low\slash negligible. The empirical findings in our case study suggest that initial implementation costs for image-based AGT are, with half the time-demand, much more gentle for EyeAutomate than for the element-based AGT framework Selenium. The maintenance cost, however, largely depends on the frequency and magnitude of changes to the UI. Greater GUI overhauls as for versions 7-15 in Fig.~\ref{fig:maintenance-cumulative} impact in this case study image-based AGT much stronger than element-based AGT, as also seen for AGT of mobile applications~\cite{Ardito2019ease}. Thus, an important lesson learned is a refinement to the previous, related studies: the stability of the GUI is the main determinant of maintenance costs and should be considered when considering them.

In practice, even test execution times may be of relevance for large test-suites, i.e., when regression testing, with developers waiting in the loop. The time it takes to execute the tests manually and automatically are both presented in Table \ref{tab:result_exec_time}. MGT is the slowest with an average of ca.~75 minutes per run of the shortlisted test-protocols $T_+$, but as mentioned before, entails no maintenance costs. EyeAutomate runs are slower in execution than Selenium tests. One caveat with EyeAutomate is that the execution, even though entirely automated, is still about half as time-consuming as the manual testing. 

\begin{table}
    \scriptsize
    \centering 
    \caption{Manual test protocol execution times compared to Selenium and EyeAutomate execution times from full-pass runs.}
    \begin{tabular}{r|ccc}
        \hline
         Exec.~Time (min.) & Manual & Selenium & EyeAutomate \\
         \hline
         Total & 75  & 7.5 & 30 \\
         \hline
         Average & 12.5 & 1.25  & 5  \\
         \hline
    \end{tabular}
    \label{tab:result_exec_time}
    \vspace{-.6cm}
\end{table}

In situations where the studied time-window using step-wise replay is too short for reaching a positive ROI, one can use statistical modeling, e.g., Bayesian data analysis to make predictions with uncertainty (like, e.g.,~\citet{afzalT08pred}). One could design a Bayesian model for the cumulative hours using version vector elements $v_i$ as predictors, as has been done in Fig.~\ref{fig:roi_prediction}. To the left of the dashed vertical line, we have the actual data collected during our case study, whereas the content to the right contains predicted values. This enables the comparison of both frameworks where, over time, the 95\% uncertainty regions (i.e., the shaded areas) increase. In other words, as we predict maintenance costs further into the future, we run into more uncertainty. 

Next, we briefly present and argue for our Bayesian model but do not have space for a more detailed introduction and description (a recent introduction to BDA in software research can be found in \cite{furia2018bayesian}). We assume the following:

\vspace{-.3cm}
{\footnotesize
\begin{IEEEeqnarray}{rCl}
\text{ch}_i & \sim & \text{Gamma-Poisson}(\lambda_i, \phi_i)\\
\log(\lambda_i) & = & \alpha + \beta_w \cdot 
v_i\\
\alpha, \beta_w & \sim & \mathcal{N}(0, 10)\\
\log(\phi) & \sim & \mathcal{\gamma}(0.5, 0.5)
\end{IEEEeqnarray}
}

\vspace{-.5cm}
In the first line, we assume that the \textit{cumulative hours (ch)}, for each framework, are distributed according to a Gamma-Poisson likelihood. The Gamma-Poisson likelihood is shaped as a negative-binomial distribution and was chosen for our model because the mean and the variance differ significantly, and its overall shape is in line with our measurements. 
In the second line, we model $\lambda$ as a linear regression where we will estimate an intercept $\alpha$ and $\beta_w$ (using a $\log$ link parameter we assume a parameter's value is the exponentiation of the linear model). In the remaining lines, we set so-called \textit{priors} for each of the parameters we want to estimate. In this case, we set very broad priors (generic weakly informative priors) according to recommended best practices~\cite{priorchoicestan}.
Note that, because versions are a mixture of days and weeks in this case study (see Fig.~\ref{fig:maintanance_cost_chronologically}), an estimation error is introduced into the model.
This does, however, not affect the estimation of the required number of hours until ROI, which can directly be derived from the y-axis of the plot.

The frequency in which manual tests are conducted highly affects the time until a positive ROI is obtained. To illustrate this, we here compare Fig.~\ref{fig:maintenance-cumulative} and Fig.~\ref{fig:roi_prediction} with weekly and monthly manual testing schemes, respectively. For weekly manual tests, the ROI is achieved within the emulated time-span of the case study (heft of the dashed line in Fig.~\ref{fig:roi_prediction}), while for monthly manual testing we must look above the horizon of our study (right of the dashed line) and extrapolate the time effort of the AGT in order to reach the time at which the intersection between the AGT and MGT models constitutes the estimated ROI. CANEA, already before the study, maintained a large set of test protocols for manual conduct. The proposed model offered them a structured, cost-effective, and exploratory way of learning about the possibilities of AGT, without fully committing to a single framework. This may not be the case for all projects, and the investment for estimation may be considered high. However, the estimation model results in executable test code, a better understanding of the testability of the SUT, as well as an update on valuable knowledge that can be applied elsewhere in a company.



Some specific lessons learned from the interviews are summarized below.
All six interviewees mentioned, to a lesser or greater extent, that programming knowledge is required to use Selenium. The developers were generally positive and found Selenium easy to use due to its familiar language and development environment, while the single tester stated that it was: '\textit{Not for me}'. This is something to consider for other companies.

The interviewees overall found the tool for creating EyeAutomate tests, EyeStudio, easy to use. The visual scripts with images made it easy to get an overview and the clickable instructions made it easy to get started. Several interviewees said that little technical knowledge was required to get started with EyeAutomate, mostly due to the outspoken \textit{intuitive workflow}. This is consistent with existing evaluations in the literature, where third-generation frameworks are perceived as more intuitive and independent of specific knowledge on programming languages or frameworks~\cite{Ardito2019ease}. However, some areas require experience, specifically when it came to handling timing issues and different so-called \textit{recognition modes}.
Also, most interviewees noted how sensitive and error-prone the image recognition felt, how the test hijacks the computer while running, and how difficult it is to reuse code.

Ultimately, the consensus of the interviewees was that an AGT can replace manual testing, to a large extent. The majority of interviewees stated that test protocols would have to be well defined up-front to be automatized. Also, exploratory testing by humans cannot be easily replaced. Automated GUI tests would further lead to more false positives and false negatives, partly because humans are more forgiving and rationale regarding the relevance of visual changes, according to the interviewees.



Another finding with potential practical bearing is the assessed generality of the model presented in Fig.~\ref{fig:roi-method}. Its construction, with slight modification, may be followed for the introduction of other manual task automation means for ROI estimation, leveraging a strong utilizing of the existing evolution of the software in form of the source-history. The GUI automation test case creation in the preparation phase is then substituted by the arbitrary automation means of choice, e.g. for build-pipeline or stress-test automation. How exactly this would play out still has to be investigated.

\subsection*{Threats to Validity}
Given that the conclusions in this paper are heavily based on a single case study, a number of limitations and threats to validity exist. Only two frameworks were compared in this study, each chosen as a representative for the respective second and third-generation of AGT frameworks. Whereas the locator principle of Selenium is generalizable among second generation AGT frameworks, does the EyeAutomate scripting syntax and GUI, as well as the image-recognition, introduce a novelty/differentiator that makes it less generalizable among other frameworks of third-generation AGT. Also, in particular for closed source third generation AGT frameworks that heavily built on image-recognition, accuracy, and thereby testing performance, may change substantially over time due to algorithmic improvements.

The test protocols were not controlled for, i.e. they were completely taken from the real-world scenario and only sampled. Version selection was done weekly, but there are many other ways to select the software versions for manual step-wise replay, as mentioned in Section \ref{method}. Also, all recorded times for manual testing, implementation, and maintenance, can be biased by the executor of the task. Thus, the ROI from one study can, at best, estimate the directionality and rough range of that of another study and project.
It can further be argued/project-specific whether bug-fixing can be considered to be part of the maintenance costs as outlined in the model description. Not committing to fixing/circumventing a bug in the version $i$ under investigation means that version $i+1$ can't be executed, so a minimal \'{}maintenance\`{} is required for the model to work. The focus here is though more on circumvention than fixing. In fact, in most cases, fixing has shown to be as hard/easy to achieve and is therefore not an \'{}external\`{} activity, but part of the maintenance in our view.
Finally, qualitative findings through the interviews are of very limited generality because of the small sample size, in particular, the involvement of a single tester.

\section{Conclusions}
\label{conclusions}

In this paper, we presented a manual step-wise replay method for ROI estimation for the introduction of automated GUI testing into existing software projects. 
The method replays the existing source-code change history with manual tester\slash developer intervention at each stage and a pre-defined process to measure ROI. 
We evaluated the method in a case study on the industrial software CANEA ONE, producing relevant insights for the company, as well as lessons learned. Differences between the two testing frameworks and approaches investigated stood out clearly. The maintenance effort per investigated version was an average of 32\% higher for EyeAutomate, but the dominating cost for both frameworks was the initial implementation time. Implementing the test-protocols for Selenium took almost twice the time than with EyeAutomate. For both frameworks, this was close to 90\% of the total cost up until reaching ROI compared to manual testing. EyeAutomate revealed, with 30 bugs, 25\% more bugs than Selenium. Also, the practical use of the two compared ATG approaches differed. Whereas Selenium requires the tester to have a programming background, EyeAutomate seems not to.

For more general conclusions, concerning test and application-specific features and characteristics, further empirical studies, covering the introduction of AGT into existing software projects, are required. The effect of the selection of versions must be further investigated. Another future direction includes the investigation of AGT test suites for systems under frequent GUI changes, i.e., how can robustness be enhanced in the AGT frameworks?
How efficient AGT frameworks are in terms of identifying bugs early, strengthening AGT's value proposition, is another relevant question. With the manual step-wise replay method introduced here, we argue that these and other important software reliability questions can be answered in a context-, company- and project-relevant way.


\bibliographystyle{IEEEtranN}
\bibliography{refs}

\end{document}